\documentclass[preprint,amsmath,amssymb,aps,showkeys,showpacs]{revtex4}
\usepackage[english]{babel}
\usepackage{graphicx}
\usepackage{graphics}
\usepackage{amsmath}
\usepackage{dcolumn}
\usepackage{amssymb}
\usepackage{bm}
\begin{document}
\title{On an atom with a magnetic quadrupole moment subject to a harmonic and a linear confining potentials}
\author{I. C. Fonseca}
\email{itallo.fisica@gmail.com}
\affiliation{Departamento de F\'isica, Universidade Federal da Para\'iba, Caixa Postal 5008, 58051-970, Jo\~ao Pessoa, PB, Brazil.}
\author{Knut Bakke}
\email{kbakke@fisica.ufpb.br}
\affiliation{Departamento de F\'isica, Universidade Federal da Para\'iba, Caixa Postal 5008, 58051-970, Jo\~ao Pessoa, PB, Brazil.}

\begin{abstract}
The quantum dynamics of an atom with a magnetic quadrupole moment that interacts with an external field subject to a harmonic and a linear confining potentials is investigated. It is shown that the interaction between the magnetic quadrupole moment and an electric field gives rise to an analogue of the Coulomb potential and, by confining this atom to a harmonic and a linear confining potentials, a quantum effect characterized by the dependence of the angular frequency on the quantum numbers of the system is obtained. In particular, it is shown that the possible values of the angular frequency associated with the ground state of the system are determined by a third-degree algebraic equation.
\end{abstract}

\maketitle

\section{Introduction}

A quantum effect that has been widely investigated in the literature associated with the electric charge, the electric dipole moment and the electric quadrupole moment is the arising of geometric quantum phases. For instance, the Aharonov-Bohm effect \cite{ab} is related to the electric charge and the He-McKellar-Wilkens effect \cite{hmw,hmw2} is related to a neutral particle with a permanent electric dipole moment. With respect to an electric quadrupole moment, in Ref. \cite{chen} is shown that the wave function of a neutral particle can acquire a geometric quantum phase that stems from the interaction between the electric quadrupole moment and a magnetic field. Moreover, in Ref. \cite{b7} is shown that a geometric quantum phase stems from the interaction between an electric field and the electric quadrupole moment, where it is analogous to the scalar Aharonov-Bohm effect for a neutral particle \cite{anan,anan2}. Other quantum effects associated with the electric charge, the electric dipole moment and the electric quadrupole moment have been investigated in the context of searching for bound states, such as the Aharonov-Bohm effect for bound states \cite{pesk,fur} and the Landau quantization \cite{landau,lin,bf25}. On the other hand, based on the magnetic multipole expansion, a quantum effect associated with geometric quantum phases for a magnetic monopole is known in the literature as the dual of the Aharonov-Bohm effect \cite{dab,dab2}. Associated with the arising of geometric quantum phases for a neutral particle with a permanent magnetic dipole moment, the corresponding quantum effect is the Aharonov-Casher effect \cite{ac}. Besides, geometric quantum phases for a neutral particle with a magnetic quadrupole moment was obtained in Ref. \cite{chen} from the interaction between the magnetic quadrupole moment and an electric field by analogy with the electric quadrupole moment case.

In particular, a quantum particle with a magnetic quadrupole moment has attracted several discussions \cite{magq1,magq2,magq3,magq,prc,chen,pra,magq4,magq5,magq6,magq7,magq8,magq10}, for instance, in atomic systems \cite{magq5,magq6}, molecules \cite{magq7,magq2}, by dealing with $P$- and $T$-odds effects in atoms \cite{magq3,magq8,magq10} and chiral anomaly \cite{magq}. In this work, we consider the single particle approximation used in Ref. \cite{prc} in order to deal with a system that consists in a moving atom with a magnetic quadrupole moment that interacts with a nonuniform electric field subject to a harmonic and a linear confining potentials. Harmonic potentials \cite{landau,greiner3,grif} have a great importance in studies of molecular structure and molecules \cite{mol,mol2,mol3,mol4}, two-dimensional quantum rings and quantum dots \cite{tan}. In the relativistic quantum mechanics context, models of the harmonic oscillator have been proposed based the Klein-Gordon and Dirac equations and became known as the Dirac oscillator \cite{osc1} and Klein-Gordon oscillator \cite{kgo}. These relativistic harmonic oscillator models have attracted the interest in studies of the Jaynes-Cummings model \cite{jay2,osc3}, quantum phase transitions \cite{extra2,extra3} and the Ramsey-interferometry effect \cite{osc6}. Besides, the interest in including a linear confining potential comes from the studies of atomic and molecular physics \cite{linear3a,linear3b,linear3c,linear3d,linear3e,linear3f}, quantum bouncer \cite{bouncer,bouncer2}, motion of a quantum particle in a uniform force field \cite{landau,balle} and relativistic quantum mechanics \cite{linear2,linear2a,linear2b,linear2c,linear2d,linear2e,linear2f,eug,scalar2,vercin,mhv}. Hence, from the classical dynamics of a moving particle with a magnetic quadrupole moment that interacts with external fields, we obtain the corresponding Schr\"odinger equation and show that the interaction between the magnetic quadrupole moment and an electric field can give rise to an analogue of the Coulomb potential. Moreover, by confining this atom to a harmonic and a linear confining potentials, we show that a quantum effect characterized by the dependence of the angular frequency on the quantum numbers of the system is obtained. As a particular case, it is shown that the possible values of the angular frequency associated with the ground state of the system are determined by a third-degree algebraic equation.

The structure of this work is as follows: in section II, we introduce the classical and quantum dynamics of a moving atom with a magnetic quadrupole moment that interacts with external fields; in section III, we discuss a particular case where the interaction between the magnetic quadrupole moment and a radial electric field can give rise to a Coulomb-type potential, and thus investigate the influence of this Coulomb-type potential on the harmonic and linear confining potentials; in section IV, we present our conclusions.

\section{Moving atom with magnetic quadrupole moment in external fields }

In this section, we start by introducing the quantum dynamics of an atom possessing a magnetic quadrupole moment which interacts with external fields. From Refs. \cite{pra,prc}, the potential energy is defined by analogy with the classical dynamics of an electric quadrupole moment in the rest frame of the particle is given by 
\begin{eqnarray}
U_{m}=-\sum_{i,j}M_{ij}\,\partial_{i}\,B_{j}, 
\label{1.1}
\end{eqnarray}
where  $\vec{B}$ is the magnetic field and $M_{ij}$ is the magnetic quadrupole moment tensor, whose characteristic is that it is a symmetric and traceless tensor \cite{pra,prc}.

Henceforth, let us consider a moving particle possessing a magnetic quadrupole moment, then, by applying the Lorentz transformation of the electromagnetic field, the magnetic field $\vec{B}$ given in Eq. (\ref{1.1}) must be replaced with $\vec{B}\rightarrow\vec{B}-\frac{1}{c^{2}}\vec{v}\times\vec{E}$ for $v\ll c$ (SI units). Thereby, the Lagrangian of this classical system is written in the form:
\begin{eqnarray}
\mathcal{L}=\frac{1}{2}m\,v^2+\vec{M}\cdot\vec{B}+\frac{1}{c^{2}}\,\vec{v}\cdot\left(\vec{M}\times\vec{E}\right),
\label{1.2}
\end{eqnarray}
where $\vec{E}$ and $\vec{B}$ are the electric and magnetic fields in the laboratory frame, respectively, and we have defined the vector $\vec{M}$ in Eq. (\ref{1.2}) in such a way that its components are determined by $M_{i}=\sum_{j}M_{ij}\,\partial_{j}$ by analogy with the vector $Q_{i}=\sum_{j}Q_{ij}\,\partial_{j}$ defined in Ref. \cite{chen}, where $Q_{ij}$ is the electric quadrupole moment tensor. From this perspective, it is simple to obtain the canonical momentum of this system: $\vec{p}=m\,\vec{v}+\frac{1}{c^{2}}\left(\vec{M}\times\vec{E}\right)$; thus, the Hamiltonian of this system is 
\begin{eqnarray}
\mathcal{H}=\frac{1}{2m}\left[\vec{p}-\frac{1}{c^{2}}(\vec{M}\times\vec{E})\right]^2-\vec{M}\cdot\vec{B}. 
\label{1.3}
\end{eqnarray}

Let us proceed with the quantization of the Hamiltonian, by replacing the canonical momentum $\vec{p}$ with the Hermitian operator $\hat{p}=-i\hbar\vec{\nabla}$. Thereby, the quantum dynamics of a moving atom with a magnetic quadrupole moment can be described by the Schr\"odinger equation:
\begin{eqnarray}
i\hbar\frac{\partial\psi}{\partial t}=\frac{1}{2m}\left[\hat{p}-\frac{1}{c^{2}}\,\vec{M}\times\vec{E}\right]^{2}\psi-\vec{M}\cdot\vec{B}\,\psi.
\label{1.4}
\end{eqnarray} 

Note that we can also extend the discussion about a moving atom with magnetic quadrupole moment in external fields by including a confining potential $V$ into the Schr\"odinger equation above. In this way, the Schr\"odinger equation becomes 
\begin{eqnarray}
i\hbar\frac{\partial\psi}{\partial t}=\frac{1}{2m}\left[\hat{p}-\frac{1}{c^{2}}\,\vec{M}\times\vec{E}\right]^{2}\psi-\vec{M}\cdot\vec{B}\,\psi+V\,\psi.
\label{1.5}
\end{eqnarray} 

Thereby, we are able to study some quantum effects of a moving atom that possesses a magnetic quadrupole moment which interacts with an electric field subject to a confining potential.

\section{confinement to a harmonic and linear confining potentials}

In this section, we show that the interaction between the magnetic quadrupole moment and a radial electric field can give rise to a Coulomb-like potential, and thus we discuss the behaviour of the system when it is subject to a harmonic and a linear confining potentials. Let us use the mathematical properties of the magnetic quadrupole tensor $M_{ij}$ and make a simple choice of the components of this tensor:
\begin{eqnarray}
M_{\rho z}=M_{z\rho}=-M,
\label{2.1}
\end{eqnarray}
where $M$ is a constant $\left(M>0\right)$ and all other components of $M_{ij}$ are null. Note that the magnetic quadrupole moment defined by the components given in Eq. (\ref{2.1}) satisfies the condition in which the magnetic quadrupole tensor $M_{ij}$ must be a symmetric and traceless matrix. These mathematical properties of the magnetic quadrupole tensor $M_{ij}$ have also been explored, for instance, in the study of geometric quantum phases, where the tensor $M_{ij}$ is considered to be diagonal \cite{chen}. Besides, let us consider an electric field given by \cite{er}
\begin{eqnarray}
\vec{E}=\frac{\lambda\,\rho}{2}\,\hat{\rho},
\label{2.2}
\end{eqnarray} 
where $\lambda=\lambda_{0}/\epsilon_{0}$, $\lambda_{0}$ is a uniform volume electric charge density, $\epsilon_{0}$ is the electric vacuum permittivity, $\rho=\sqrt{x^{2}+y^{2}}$ and $\hat{\rho}$ is an unit vector in the radial direction. The field configuration given in Eq. (\ref{2.2}) was explored in studies of the Landau quantization for neutral particles with a permanent magnetic dipole moment \cite{er}. Note that the Landau quantization \cite{landau} is characterized by the presence of a uniform magnetic field on the path of a charged particle given by a vector potential $\vec{A}=\frac{B\rho}{2}\,\hat{\varphi}$, where $B$ is the intensity of the magnetic field and $\hat{\varphi}$ is a unit vector in the azimuthal direction. In the present case, the effective vector potential given in Eq. (\ref{1.5}) becomes $\vec{A}_{\mathrm{eff}}=\vec{M}\times\vec{E}=-\frac{M\,\lambda}{2}\,\hat{\varphi}$. Despite the presence of the effective vector potential cannot yield the Landau quantization for a moving atom with a magnetic quadrupole moment, the importance of it is that it contributes to the arising of an analogue of the Coulomb potential as we shall see in the following.

Finally, let us confine the system described by Eqs. (\ref{2.1}) and (\ref{2.2}) to a harmonic and a linear confining potentials, which can be introduced into the Schr\"odinger equation (\ref{1.5}) through
\begin{eqnarray}
V\left(\rho\right)=\frac{1}{2}\,m\,\omega^{2}\,\rho^{2}+\eta\,\rho.
\label{2.3}
\end{eqnarray} 
In particular, the linear scalar potential given in Eq. (\ref{2.3}) has been proposed to describe the confinement of quarks \cite{linear,linear1} due to experimental data show a behaviour of the confinement to be proportional to the distance between the quarks \cite{linear4,linear4a,linear4b,linear4c}. It has also been explored in studies of the quark-antiquark interaction as a problem of a relativistic spinless particle which possesses a position-dependent mass, where the mass term acquires a contribution given by a interaction potential that consists of a linear and a harmonic confining potential plus a Coulomb potential term \cite{bah}. Besides, the linear scalar potential given in Eq. (\ref{2.3}) has attracted a great interest in atomic and molecular physics as pointed out in Refs. \cite{linear3a,linear3b,linear3c,linear3d,linear3e,linear3f} and in several discussions of relativistic quantum mechanics \cite{linear2,linear2a,linear2b,linear2c,linear2d,linear2e,linear2f,eug,scalar2,vercin,mhv}.

We simplify our calculations by working with the units $\hbar=c=1$, therefore the Schr\"odinger equation (\ref{1.5}) becomes
\begin{eqnarray}
i\frac{\partial\psi}{\partial t}&=&-\frac{1}{2m}\left[\frac{\partial^{2}}{\partial\rho^{2}}+\frac{1}{\rho}\frac{\partial}{\partial\rho}+\frac{1}{\rho^{2}}\,\frac{\partial^{2}}{\partial\varphi^{2}}+\frac{\partial^{2}}{\partial z^{2}}\right]\psi-i\,\frac{M\lambda}{2m\rho}\,\frac{\partial\psi}{\partial\varphi}+\frac{M^{2}\lambda^{2}}{8m}\,\psi\nonumber\\
[-2mm]\label{1.7}\\[-2mm]
&+&\frac{1}{2}\,m\,\omega^{2}\,\rho^{2}\,\psi+\eta\,\rho\,\psi.\nonumber
\end{eqnarray}

We can see that the quantum operators $\hat{L}_{z}=-i\partial_{\varphi}$ and $\hat{p}_{z}=-i\partial_{z}$ commute with the Hamiltonian operator given in the right-hand side of (\ref{1.7}), then, a particular solution to Eq. (\ref{1.7}) can be written in terms of the eigenvalues of the operator $\hat{p}_{z}$, and $\hat{L}_{z}$: 
\begin{eqnarray}
\psi\left(t,\rho,\varphi,z\right)=e^{-i\mathcal{E}t}\,e^{i\,l\,\varphi}\,e^{ikz}\,R\left(\rho\right),
\label{1.8}
\end{eqnarray}
where $l=0,\pm1,\pm2,\ldots$, $k$ is a constant, and $R\left(\rho\right)$ is a function of the radial coordinate. Thereby, substituting the solution (\ref{1.8}) into Eq. (\ref{1.7}), we obtain
\begin{eqnarray}
R''+\frac{1}{\rho}R'-\frac{l^{2}}{\rho^{2}}R-\frac{M\lambda\,l}{\rho}\,R-m^{2}\omega^{2}\rho^{2}\,R-2m\,\eta\,\rho\,R+\zeta^{2}\,R=0,
\label{1.9}
\end{eqnarray}
where we have defined the following parameter:
\begin{eqnarray}
\zeta^{2}&=&2m\mathcal{E}-k^{2}-\frac{M^{2}\lambda^{2}}{4}.
\label{1.10}
\end{eqnarray}

Observe that the fourth term on the left-hand side of Eq. (\ref{1.9}) plays the role of a Coulomb-like potential for $l\neq0$. This term stems from the interaction between the electric field (\ref{2.2}) and the magnetic quadrupole moment given in Eq. (\ref{2.1}) due to the presence of the term proportional to $\vec{A}_{\mathrm{eff}}\cdot\hat{p}=\left(\vec{M}\times\vec{E}\right)\cdot\hat{p}$ in the Schr\"odinger equation (\ref{1.7}). Note that for $l=0$, this term vanishes and there is no presence of the interaction between the electric field and the magnetic quadrupole moment, which makes no sense for our discussion. In this way, the focus of our discussion is for $l\neq0$. Thereby, let us make a change of variable given by $\xi=\sqrt{m\omega}\,\rho$, thus, the radial equation (\ref{1.9}) becomes
\begin{eqnarray}
R''+\frac{1}{\xi}\,R'-\frac{l^{2}}{\xi^{2}}\,R-\frac{\delta}{\xi}\,R-\xi^{2}\,R-\alpha\,\xi\,R+\frac{\zeta^{2}}{m\omega}\,R=0,
\label{1.11}
\end{eqnarray}
where the parameters $\delta$ and $\alpha$ are defined as
\begin{eqnarray}
\delta=\frac{M\,\lambda\,l}{\sqrt{m\omega}};\,\,\,\,\,\,\,\alpha=\frac{2\,m\,\eta}{\left(m\omega\right)^{3/2}}.
\label{1.12}
\end{eqnarray}

By analysing the asymptotic behaviour of the possible solutions to Eq. (\ref{1.11}), we have that the asymptotic behaviour is determined for $\xi\rightarrow0$ and $\xi\rightarrow\infty$ \cite{heun,eug,mhv,vercin}. This allows us to write the function $R\left(\xi\right)$ in terms of an unknown function $H\left(\xi\right)$ in such a way that it is a regular function at the origin:
\begin{eqnarray}
R\left(\xi\right)=e^{-\frac{1}{2}\,\xi^{2}}\,e^{-\frac{\alpha}{2}\,\xi}\,\xi^{\left|l\right|}\,H\left(\xi\right).
\label{1.13}
\end{eqnarray}
Substituting (\ref{1.13}) into (\ref{1.11}), we obtain the following equation:
\begin{eqnarray}
H''+\left[\frac{\theta}{\xi}-\alpha-2\xi\right]H'+\left[g-\frac{\left(\theta\,\alpha+2\delta\right)}{2\xi}\right]H=0,
\label{1.14}
\end{eqnarray}
where 
\begin{eqnarray}
g=\frac{\zeta^{2}}{m\omega}+\frac{\alpha^{2}}{4}-2-2\left|l\right|;\,\,\,\,\,\,\,\theta=2\left|l\right|+1.
\label{1.15}
\end{eqnarray}
The function $H\left(\xi\right)$ is a solution to the second order differential equation (\ref{1.14}) and it is known as the biconfluent Heun function \cite{heun,eug,bm,b50,b50a}:
\begin{eqnarray}
H\left(\xi\right)=H\left(2\left|l\right|,\,\alpha,\,\frac{\zeta^{2}}{m\omega}+\frac{\alpha^{2}}{4},\,2\delta,\,\xi\right).
\label{1.16}
\end{eqnarray} 

Our goal is to obtain bound states solutions, then, let us use the Frobenius method \cite{arf,eug,b50,b50a}. Thereby, the solution to Eq. (\ref{1.14}) can be written as a power series expansion around the origin:
\begin{eqnarray}
H\left(\xi\right)=\sum_{j=0}^{\infty}\,c_{j}\,\xi^{j}.
\label{1.17}
\end{eqnarray} 
Therefore, substituting the series (\ref{1.17}) into (\ref{1.14}), we obtain the recurrence relation:
\begin{eqnarray}
c_{j+2}=\frac{\left[2\alpha\left(j+1\right)+\theta\alpha+2\delta\right]}{2\left(j+2\right)\,\left(j+1+\theta\right)}\,c_{j+1}-\frac{\left(g-2j\right)}{\left(j+2\right)\,\left(j+1+\theta\right)}\,c_{j}.
\label{1.18}
\end{eqnarray}

By starting with $c_{0}=1$ and using the recurrence relation (\ref{1.18}), we can calculate other coefficients of the power series expansion (\ref{1.17}). As an example, the terms $c_{1}$ and $c_{2}$ are
\begin{eqnarray}
c_{1}&=&\frac{\alpha}{2}+\frac{\delta}{\theta};\nonumber\\
[-2mm]\label{1.19}\\[-2mm]
c_{2}&=&\frac{\left[2\alpha+\theta\alpha+2\delta\right]}{4\left(1+\theta\right)}\left(\frac{\alpha}{2}+\frac{\delta}{\theta}\right)-\frac{g}{2\left(1+\theta\right)}.\nonumber
\end{eqnarray}

Bound state solutions can be obtained by imposing that the power series expansion (\ref{1.17}) or the biconfluent Heun series becomes a polynomial of degree $n$. From the recurrence relation given in Eq. (\ref{1.18}), we have that the power series expansion (\ref{1.17}) becomes a polynomial of degree $n$ by imposing two conditions \cite{bm,eug,b50,b50a}:
\begin{eqnarray}
g=2n\,\,\,\,\,\,\mathrm{and}\,\,\,\,\,\,c_{n+1}=0,
\label{1.20}
\end{eqnarray}
where $n=1,2,3,\ldots$. By using the expression of $g$ given in Eq. (\ref{1.15}), then, the condition $g=2n$ yields
\begin{eqnarray}
\mathcal{E}_{n,\,l}=\omega\left[n+\left|l\right|+1\right]-\frac{\eta^{2}}{2m\,\omega^{2}}+\frac{M^{2}\lambda^{2}}{8m}+\frac{k^{2}}{2m}.
\label{1.21}
\end{eqnarray}

Equation (\ref{1.21}) corresponds to the spectrum of energy of a moving atom with a magnetic quadrupole moment subject to a harmonic potential and a linear confining potential under the influence of an analogue of the Coulomb potential. As we have seen, the Coulomb-type potential stems from the interaction between the magnetic quadrupole moment defined in Eq. (\ref{2.1}) and the radial electric field given in Eq. (\ref{2.2}). As discussed previously, the Coulomb-type potential is defined for all values of the quantum number $l$ that differ from zero, therefore the energy levels (\ref{1.21}) are defined for $l\neq0$, otherwise there is no presence of the Coulomb-type potential.

However, we have not analysed the condition $c_{n+1}=0$ imposed in Eq. (\ref{1.20}). For this purpose, let us assume that the angular frequency $\omega$ can be adjusted in such a way that the condition $c_{n+1}=0$ is satisfied. With this assumption, we have that both conditions imposed in Eq. (\ref{1.20}) are satisfied and we obtain a polynomial solution to the function $H\left(\xi\right)$ given in Eq. (\ref{1.17}). As a consequence, we obtain an expression involving the angular frequency and the quantum numbers $\left\{n,\,l\right\}$ of the system, whose meaning is that only specific values of the angular frequency $\omega$ are allowed and depend on the quantum numbers $\left\{n,\,l\right\}$. Thereby, we label 
\begin{eqnarray}
\omega=\omega_{n,\,l}.
\label{1.22}
\end{eqnarray}

This corresponds to a quantum effect characterized by a dependence of the angular frequency of the harmonic potential on the quantum numbers $\left\{n,\,l\right\}$ of the system that stems from the influence of the Coulomb-type potential on the harmonic and linear confining potentials. From the mathematical
point of view, this relation involving the angular frequency of the harmonic oscillator on the quantum numbers $\left\{n,\,l\right\}$ results from the fact that the exact solutions to Eq. (\ref{1.16}) are achieved for specific values of harmonic oscillator frequency. In recent years, analogue effects of this angular frequency dependence on the quantum numbers of the system have been investigated in different quantum mechanical contexts \cite{bm,eug,b50,b50a,bf40}.

As an example, let us consider $n=1$, which corresponds to the ground state, and analyse the condition $c_{n+1}=0$. For $n=1$, we have $c_{2}=0$. The condition $c_{2}=0$ imposes that the angular frequency $\omega_{1,\,l}$ satisfies the third-degree algebraic equation \cite{b50,b50a,eug}: 
\begin{eqnarray}
\omega_{1,\,l}^{3}-\frac{\left(M\,\lambda\,l\right)^{2}}{2m\theta}\,\omega_{1,\,l}^{2}-\frac{\eta\,M\,\lambda\,l}{m\theta}\left(1+\theta\right)\,\omega_{1,\,l}-\frac{\left(2+\theta\right)\,\eta^{2}}{2m}=0.
\label{1.23}
\end{eqnarray}

Despite Eq. (\ref{1.23}) having at least one real solution, we do not write it because its expression is very long. Moreover, for $n=1$ we have the simplest case of the function $H\left(\xi\right)$ which corresponds to a polynomial of first degree:
\begin{eqnarray}
H_{1,\,l}\left(\xi\right)=1+\left(\frac{m\,\eta}{\left(m\,\omega_{1,\,l}\right)^{3/2}}+\frac{M\,\lambda\,l}{\theta\,\left(m\,\omega_{1,\,l}\right)^{1/2}}\right)\,\xi.
\label{1.24}
\end{eqnarray}

In this way, the general expression for the energy levels (\ref{1.21}) is given by:
\begin{eqnarray}
\mathcal{E}_{n,\,l}=\omega_{n,\,l}\left[n+\left|l\right|+1\right]-\frac{\eta^{2}}{2m\,\omega_{n,\,l}^{2}}+\frac{M^{2}\lambda^{2}}{8m}+\frac{k^{2}}{2m}.
\label{1.25}
\end{eqnarray}

Hence, we have obtained bound state solutions for a moving atom with a magnetic quadrupole moment subject to harmonic and linear confining potentials under the influence of both attractive or repulsive Coulomb-type potentials. Finally, we have seen that the influence of the Coulomb-type potential on the harmonic and the linear confining potentials gives rise to a quantum effect characterized by the dependence of the angular frequency of the harmonic potential on the quantum numbers $\left\{n,\,l\right\}$, whose meaning is that only specific values of the angular frequency of the harmonic potential are allowed in the system in order that a polynomial solution to the function $H\left(\xi\right)$ can be obtained \cite{eug,bm,b50,b50a}.

\section{conclusions}

We have investigated the influence of Coulomb-type potential that stems from the interaction between a radial electric field and the magnetic quadrupole moment of a atom on the harmonic and linear confining potential. We have seen that bound state solutions of the Schr\"odinger equation can be obtained for both attractive or repulsive Coulomb-type potentials. Besides, the influence of Coulomb-type potential on the linear confining potential and the harmonic potential has yielded a quantum effect characterized by the dependence of the angular frequency on the quantum numbers of the system. In particular, we have shown that the possible values of the angular frequency associated with the ground state of the system are determined by a third-degree algebraic equation.

It is worth observing that we have considered relativistic corrections of the field up to $O\left(\frac{v^{2}}{c^{2}}\right)$. An interesting case would be the analysis of this system under the influence of the relativistic corrections which includes terms of order $\mathcal{O}\left(\frac{v^{2}}{c^{2}}\right)$. In this case, the relativistic effects can give rise to an effective mass \cite{whw} in which can be interesting in studies of position-dependent mass systems \cite{pdm,pdm2,pdm3,pdm4}. We hope to bring these discussions in the near future.

\acknowledgments

The authors would like to thank the Brazilian agencies CNPq and CAPES for financial support.

\end{document}